# High efficiency resonant-metasurface germanium photodetector with ultra-thin intrinsic layer

Jinwen Song[1,#], Shuai Yuan[1,#], Jinsong Xia[1,*]

[1]Wuhan National Laboratory for Optoelectronics, Huazhong University of Science and Technology, Wuhan 430074, China

*Corresponding author: jsxia@hust.edu.cn

#These authors contributed equally to this work

**Photodetectors at telecommunications-band with high efficiency and high speed are becoming increasingly important as the booming of big data, 5G, internet of things, cloud computing, artificial intelligence and relevent applications. Silicon-based Germanium photodetectors exhibit great potential in reducing the cost and power dissipation, due to its compatibility of monolithic integration with signal-processing electronics. We report the first demonstration of normal incident resonant-metasurface germanium photodetector, to address the trade-off between quantum efficiency, bandwidth and wavelength coverage for free-space detectors. With an ultra-thin intrinsic layer thickness of 350 nm, a high external quantum efficiency of more than 60% and clearly open eyes at the speed of 20 Gbps are achieved, for a 30 μm-diameter device. The photodetector employs multiple trapped-mode resonances to enhance the localized electromagnetic field, which not only enhances the external quantum efficiency by more than 300% at 1550 nm, but also extends to the whole C band. Further simulation and measurement with small incident spot size show the feasibility to achieve >50 GHz 3 dB bandwidth by simply reducing the mesa size, with minor sacrifice of the enhanced absorption. This work paves the way for the future development of low-cost, high efficiency normal incident germanium photodetectors operating at data rates of 50 Gbps or higher.**

**Introduction:**

Since the birth of the first fiber optic communication system, the world has witnessed its continuous development with ever-increasing transmission rate and capacity. The demand of high quality and low latency communications has been driving the development of high speed optical and electric chips. To date, the power and heat dissipation turns out to be another important factor, due to the high assembly density of communication equipment. Monolithic integration of optical devices and electric circuits on silicon wafer holds the key to fulfill all these demands at the same time. Which requires that the material and fabrication processes of the optical components should be compatible with the silicon complementary metal–oxide–semiconductor (CMOS) circuits. The indirect band gap of group IV elements block its application in semiconductor lasers, while the silicon[1] and silicon-based Ge photodetectors[2,3] have been developed for different kinds of applications. The integration of these photodiodes with the necessary driving electric circuit and trans-impedance amplifier on the same wafer significantly reduces the cost of packaging the transceiver sub-assembly and improves performance, as well as the cost per gigabit per second.

Currently, photodetector used in optical transceiver is dominated by III-V compound semiconductor, as its constituent can be adjusted flexibly to satisfy the applications in different communication bands, despite the CMOS incompatibility. In comparison, the use of CMOS compatible Ge photodetectors is quite limited due to the weak absorption efficiency at C band. For normal incidence Ge photodetectors in high speed communication, the carrier transit time is so critical that the intrinsic layer of the photodetector needs to be thin enough. For example, to guarantee > 50 GHz 3-dB bandwidth in a Ge photodetector, the thickness of the intrinsic-Ge layer should be thinner than 500 nm. Considering the absorption efficiency of Ge at 1550 nm, which is around 7000 cm$^{-1}$ (ref. 7), the responsivity of the Ge photodetector with a 500-nm-thick intrinsic layer will be less than 0.3 A/W. Waveguide photodetector can address the trade-off between quantum efficiency and bandwidth, by separating the photon-transmission direction from the carrier-transit direction. However, the strict coupling method of waveguide photodetector add the cost in packaging and limits its application in cost efficient optical transceiver.

To overcome this difficulty in normal-incidence Ge phot-odetector, different kinds of method have been developed without increasing the thickness of intrinsic layer. A method to enhance the absorption in a wide spectra is the incorporation of Sn in the Ge crystal matrix. The indirect and direct bandgap of the mixed GeSn crystal shift to longer wavelengths than Ge, resulting in larger absorption efficiency at C band. However, the dark current of the GeSn photodetector increases with the increasing Sn content, due to the reduction of the GeSn bandgap and the increase of defect density of the film[5]. Such a high dark current will severely limit the sensitivity of the photodetector, which is of key importance in fiber optical communication. Apart from the change of material, different kinds of structures can be applied to enhance the photon response. The Fabry-Perot cavity is a traditional way to enhance the light-matter interactions by trapping light inside it. A number of resonant cavity enhanced (RCE) photodetectors have been demonstrated on Ge-on-Si platform, employing distributed Bragg reflectors or reflection gratings. Although enhanced responsivity has been achieved in such devices, they are usually wavelength sensitive. The trade-off between absorption enhancement and wavelength coverage has limited their applications in WDM optical communications. Very recently, a new kind of photon-trapping Si photodetectors was demonstrated[1,13] for wavelength around 850 nm. Two dimentional micro- and nanoscale structures are introduced to diffract the vertically incident light into a horizontal waveguide mode, and thus the light is trapped in the intrinsic layer by total internal reflection to enhance the abosrption efficiency. Multifold enhancement of responsivity was achieved (refs 10,13), showing great potential in optical communications. Further demonstration of Ge photodetector with similar structures[12] increases the responsivity from 0.6 A/W to 0.91 A/W with an intrinsic-layer thickness of 2 μm, at the wavelength of 1550 nm (ref. 9). However, despite the high responsivity and wide wavelength coverage, the further high speed performance of these photodetectors are limited by the thick intrinsic layer.

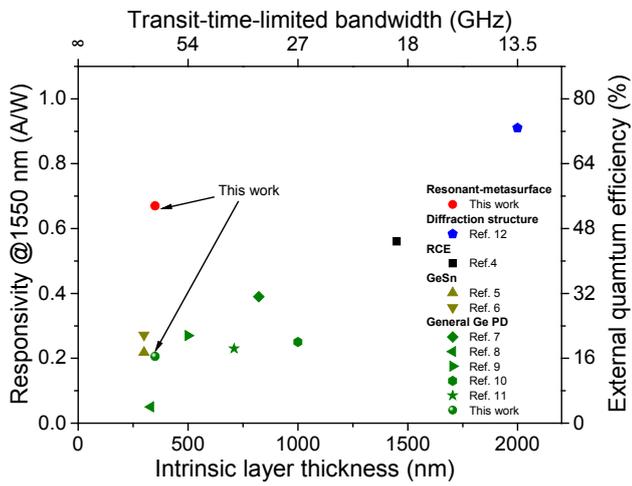

**Figure 1 | Summary of published responsivity (@1550 nm) with respect to the intrinsic layer thickness and transit-time-limited bandwidth for surface-illuminated pin germanium photodetector.** The black symbol is measured for resonant cavity enhanced (RCE) Ge PD[4], the dark yellow symbols for GeSn PDs[5,6], the green symbols for general Ge PDs of publications[7-11] and this work, the blue symbols for Ge PD integrated with diffraction structure[12], and the resonant-metasurface Ge PD presented in this paper (red circle).

Up to date, no demonstrations has been given, for normal-incidence Si-based Ge photodetector with a thin intrinsic layer, to achieve multifold enhancement responsivity in the whole C band without sacrificing other performance of the device. A summary of experimentally measured responsivities with respect to intrinsic layer thickness for different Si-based Ge photodetectors is presented in Figure 1. The trade-off between responsivity and transit-time-limited bandwidth can be clearly seen for the published results. Here we demonstrate a surface-illuminated Ge-on-Si pin photodetector with a 350-nm-thick intrinsic layer, which employs resonant-metasurface to enhance the light-matter interactions. A peak value of 0.713 A/W is achieved at 1540 nm, corresponding to an external quantum efficiency as high as 62%. In addition, the metasurface structure is specially designed to excite multiple resonances to broaden the wavelength coverage of optical resonance. As a consequence, responsivity above 0.5 A/W was achieved in the wavelength range from 1500 nm to 1570 nm. Eye diagrams of the fabricated device clearly open at 20 Gbps. A large 3 dB bandwidth of more than 50 GHz is expectable by simply reduce the mesa diameter to 10 μm with minor sacrifice of the enhanced responsivity.

**Design and simulations**

As is well-known, resonance is one of the most common way to enhance light-matter interactions[14-18], as it increases photon lifetime and local field intensity. The combination of resonant structure and photodetector can bring with stronger absorption by trapping light inside[19,20]. However, one key characteristic of resonance is wavelength selectivity, and a wavelength selective photodetector is unsuitable in wavelength division multiplex (WDM) optical communication systems. Employing multiple resonances to broaden wavelength coverage is a method worth trying, thus a resonant structure which can support multiple resonance modes overlapping with each other is in urgent need.

Recently, the manipulation of optically induced Mie reso-nances in low-loss all-dielectric metasurface[21-30] has become one of the fastest growing branches of nanophotonics. The coexistence of strong electric and magnetic resonances in dielectric metasurfaces brings flexible and diverse designs for different applications. The collective resonance of every single unit cell in metasurface can strongly enhance the light-material interaction. Typically, the resonances in dielectric metasurface are consist of different orders of electric and magnetic multipole oscillations, which can be further divided into pairs of in-phase and anti-phase modes. In addition, the resonant wavelength of different modes can be conveniently adjusted by simply changing the basic structural parameters of the dielectric metasurface. This characteristics makes resonant-metasurface a perfect candidate to address the weak absorption in Ge at C band.

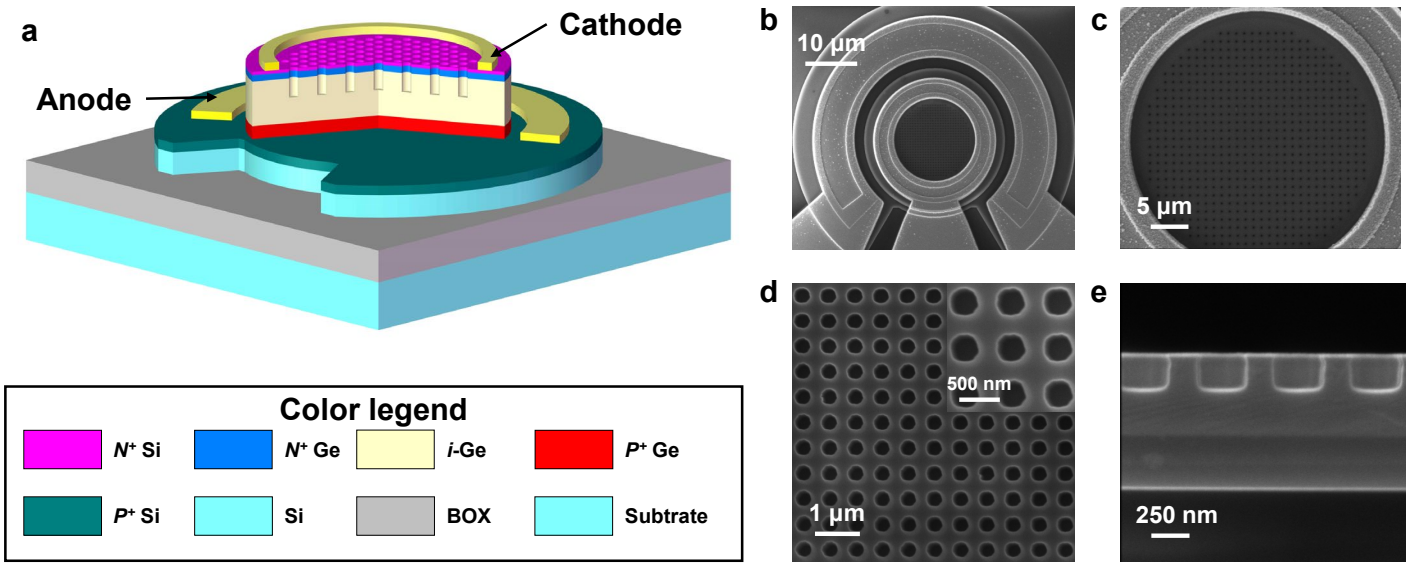

**Figure 2 | Germanium photodetector with resonant-metasurface enhancing light absorption. a**, Schematic of the resonant-metasurface germanium photodetector with an ultrathin absorbing layer. The materials of different colored layers from top to bottom: n-Si layer, pink; n-Ge, sapphire blue; i-Ge, pale yellow; p-Ge, red; p-Si, green black; top Si of SOI, cyan; buried oxide layer (BOX) of SOI, gray; Si substrate of SOI, cyan. **b,c**, Scannin electron micrograph (SEM) image (**b**) and enlarged image of the active region (**c**) of a resonant-metasurface germanium photodetector. **d**, resonant-metasurface intergrated in the photodetectors. Inset: Enlarge image of nine unit cells. **e**, A cross section of resonant-metasurface etched into i-Ge layer of the photodetector.

The proposed resonant-metasurface Ge photodetector (RM-Ge-PD) is a typical mesa-type PIN photodetector based on silicon-on-insulator platform. Details of the structure are schematically shown in Fig. 2a. Thickness of the top Si layer and buried $SiO_2$ layer of the basic SOI wafer are 220 nm and 2000 nm, respectively. A 100 nm Si buffer layer is grown using MBE, followed by a 100 nm p-Ge layer, a 350 nm i-Ge layer, a 50 nm n-Ge layer, and a 25 nm n-Si layer. The ultra-thin absorption region is designed to comprise a 350-nm-thick i-Ge layer to minimize the transit time for both electrons and holes. High doping in the p and n layers decreases the minority carrier lifetime and minimizes the diffusion of the photocarriers generated in the p and n layers into the high field i-region, as well as reducing the series resistance. The designed resonant-metasurface is a two-dimensional periodic structure. Sub-wavelength scale holes are shallowly etched into the mesa surface, with depth of around 215 nm. The holes are arranged as square lattice, and are fabricated as uniform cylindrical shape. At last, a 300 nm PECVD $SiO_2$ layer is deposited on top layer to passivate and protect the device. The much higher refractive indexes of Si and Ge than $SiO_2$ provides the necessary field confinement for the sustainment of resonant modes. For comparison, a portion of the devices on the same wafer have no pattern on the active mesa surface, while the others are fabricated with different parameters of metasurface. The detailed fabrication processes is available in Supplementary Information.

Figs. 2b-2e show the scanning electron microscope (SEM) images for a 30-μm-diameter RM-Ge-PD. It is important to notice that the top passivation layer of $SiO_2$ is wet etched for a clear view of the metasurface structure in Fig. 2d and Fig. 2e. As we can see, the periodic metasurface structure are fabricated with great uniformity, both in shape and etching depth. Similar images with $SiO_2$ on top for both top view and cross-setion are available in Supplementary Information.

To explore the resonant characteristic of each mode, a primary simulation is conducted on an infinite two dimensional metasurface, using the finite-difference time-domain (FDTD) method. Periodic boundary condition is applied and the two-dimensional holes are arranged in x-y

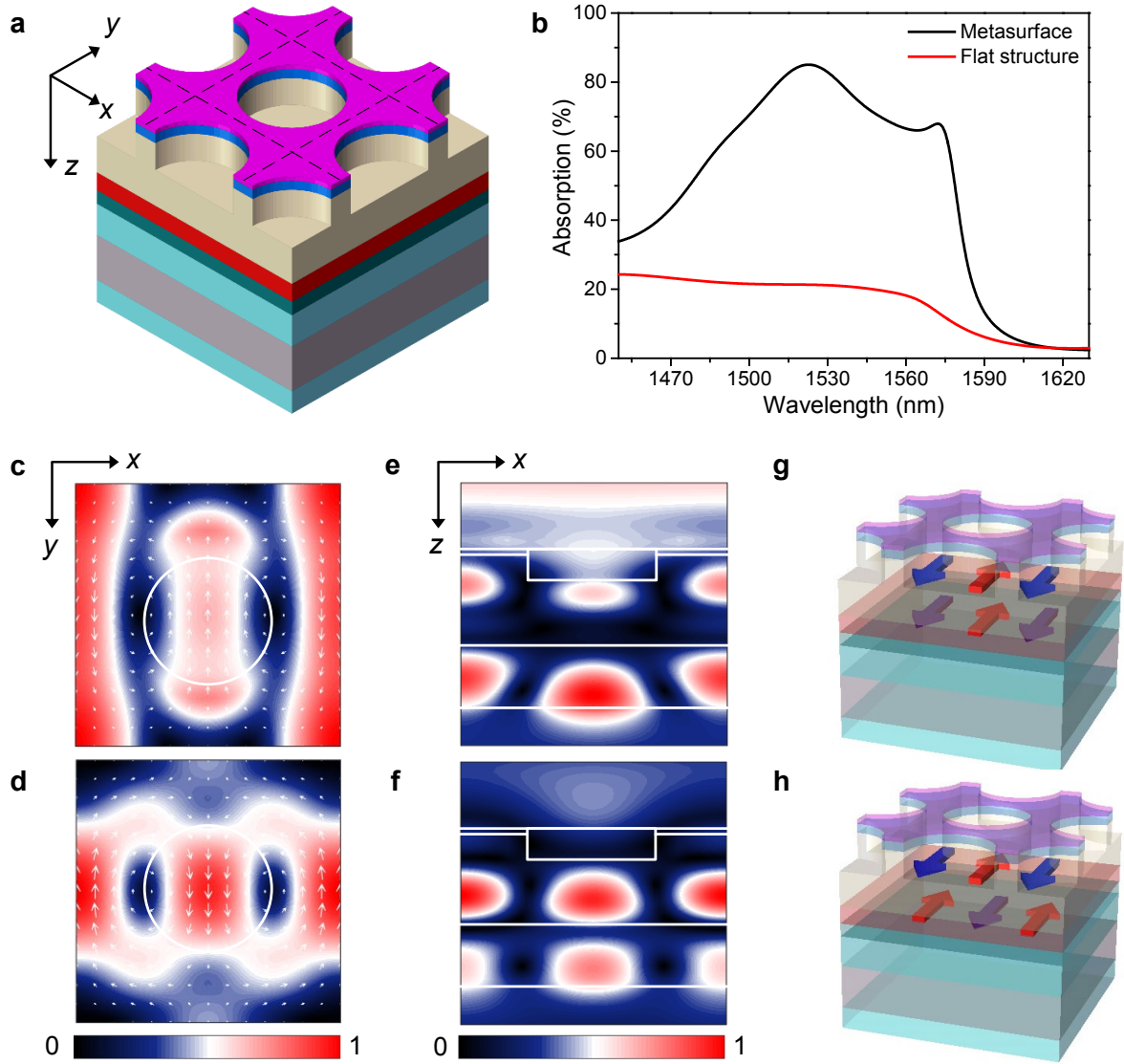

**Figure 3 | FDTD simulation of infinite periodic resonant metasurface. a**, One unit cell of the periodic metasurface. **b**, Calculated absorption spectra of the metasurface and flat structure. **c,d**, normalized electric field intensity distributions and vector distributions of mode $M_{E0}$ (**c**) and mode $M_{E1}$ (**d**) are given in the x-y plane at the center of the intrinsic layer, the wide circles depict the position of the etched hole. **e,f**, The longitudinal section of the normalized electric field intensity distributions of mode $M_{E0}$ (**e**) and mode $M_{E1}$ (**f**) are given in the x-z plane at the center of the structure. **g,h**, Schematics of the multipole resonances for mode $M_{E0}$ (**g**) and mode $M_{E1}$ (**h**) show the relationship between them.

plane, as shown in Fig. 3a. The hole-array is given the same period in x and y direction for polarization independency. Period, etching depth and radius of the holes, are properly chosen to achieve multiple resonant modes overlapping with each other at C band. The permittivities of each material used in FDTD simulations are experimentally measured using spectroscopic ellipsometer. Under normal incidence of plane wave, the structural asymmetry brought by the hole-array gives rise to collective resonance of every unit cell, thus the photons are trapped inside and gradually absorbed in Ge. The calculated absorption spectra of the resonant metasurface and the referential structure with no metasurface are given in Fig. 3b. A strong absorption enhancement is clear in the wavelength range from 1480 nm to 1580 nm, with two distinct resonance peaks at 1506 nm and 1558 nm. Figs. 3c and 3d show the electric field intensity distributions and electric field vector distributions of these two resonant modes at horizontal cross-section in x-y plane, respectively. It is clear that these two resonances are both electric modes with pairs of anti-phased electric dipole oscillations, we mark them as $M_{E0}$ and $M_{E1}$. Furthermore, from the longitudinal section electric field intensity distributions in the x-z plane, shown in Figs. 3e and 3f, these two resonance modes are of high order with two maximums in the z direction. Figs. 3g and 3h schematically depict the phase of the electric dipoles in the two resonant modes, respectively. The blue and red arrows refer to dipole oscillations with opposite phase. In this way, modes $M_{E0}$ and

$M_{E1}$ are the same order electric multipole resonances with opposite phase. It can be seen that most of the electric field is confined in the Ge and Si layers, because of their high refractive index compared with the SiO$_2$ BOX layer and cladding.

From the point of view of dielectric metasurface, the origin of these resonances is the structural asymmetry introduced by the hole-array. The asymmetry between the center and edge of the unit cell gives rise to incomplete destructive interference between adjacent dipoles with opposite phase, resulting in a smaller net dipole moment. The quality factor of the resonant mode is determined by the material loss and the resonant coupling to free space. In consideration of absorption enhancement and spectral width, the size of the hole must be properly chosen to achieve a Q-factor neither too large nor too small. Furthermore, a good overlap between the electromagnetic field and the Ge-Si layer is beneficial for the light absorption in the intrinsic layer. Compared with the absorption in the unpatterned multi-layer structure, the absorption in the metasurface structure is enhanced by over 4 times at the resonance peaks. This enhancement is mainly attributed to the slow light effect and enhanced local field intensity brought by the collective resonance. One thing worth noticing is that, although there are only two distinct resonant peaks in the calculated absorption spectrum, there are actually four resonant modes in the wavelength range from 1480 nm to 1580 nm. Besides the two electric resonant modes discussed above, there are two magnetic counterparts around the wavelength of 1485 nm and 1550 nm, respectively. A detailed discussion of these four resonant modes is given in Supplementary Information.

**Results**

In addition to the periodic simulation, a further 3D-FDTD calculation is performed on a full structure of the Ge photodetector, as schematically depicted in Fig. 4a. Size of the mesa is set 15 μm, the same as the experimentally fabricated device. Pattern of the hole-array is exactly the same as experiment, radius and etching depth are extracted from the SEM results. Calculated absorption spectra of the full mesa structure are shown in Fig. 4b, the absorption efficiencies are greatly enhanced compared with the flat PD. While the results are a little different from that of the infinite periodic hole-array, as the incident source is changed to Gaussian beam with a spot diameter of 10 μm, and the number of period is limited by the size of the mesa. Thus the effective region of the metasurface is limited and the strength of resonances are weakened, resulting in a smaller peak absorption efficiency than the case of infinite period. Considering the large difference of refractive index between the SiO$_2$ box layer and the adjacent Si, strong reflection happens at the interface. The wavelength of the reflected light, given the index of Si and Ge, is beneath the period of metasuface when the wave hits the hole-array. In this situation, the hole-array can acts as a two-dimensional grating coupler which couples the vertically reflected light into the horizontal wave guide. The diffraction of the reflected light can be clearly seen in the dynamic variation of the electric field in FDTD simulation. Despite the low diffraction efficiency and that the incident light is almost absorbed in the resonance of metasurface, this diffraction effect still helps to enhance the absorption efficiency and broaden the spectra width. Details of the diffraction simulation is available in Supplementary Information.

The experimentally measured EQEs of the patterned and unpatterned Ge photodetectors are also given in Fig. 4b, using a tunable laser with operation wavelength range from 1500 to 1630 nm. A splendid peak EQE value of 62% is achieve at 1530 nm, with strongly enhanced absorption from 1500 nm to 1545 nm (EQE > 50%). Compared with the simulation results, the measured EQEs of the RM-Ge-PD is lower to some extent, as the simulated value is around 80% at 1530 nm. The main reason for this discrepancy is the doped Ge layers, where the photo-generated carriers can hardly contribute to the photocurrents, due to the weak electric field in the doped regions. The thickness of the intrinsic layer is 350 nm out of the total 500 nm Ge, which means a considerable part of the incident photon will be dissipated in the doped regions. In addition, the absorption is also related to the electromagnetic field distribution of the resonant modes. Overlap between the field with the doped region will severely degenerate the EQE performance, which is clear from the difference between the two resonant modes. The gap of EQE between simulation and experiment at mode $ME_1$ is much larger than that of mode $ME_0$, as the overlap with p-doped Ge region is larger for mode $ME_1$ from the distributions in Fig. 3.

Figs. 4c and 4d give the responsivity values and the corresponding enhancement values of the experimentally fabricated RM-Ge-PDs, with different periods of the metasurface hole-array, respectively. Resonant wavelength of the metasurface is determined by the structural parameters. The peak wavelength red-shifts monotonously as the metasurface period increases, with peak EQE enhancement values around 350%. This characteristic of the metasurface gives great flexibility for the design of RM-Ge-PD working in different wavelength range. For example, a high EQE value of more than 0.7 A/W is achieved under period of 540 nm at λ = 1540 nm, due to the strong resonance at this wavelength. For the period of 550 nm, the resonance moves to longer wavelength region where the absorption efficiency of Ge is smaller. The peak EQE values decreases while the wavelength coverage is broadened to >0.5 A/W for λ = 1500-1570 nm. On the other hand, the results also implies that the diffraction effect of the back reflected wave is minor in our structure, as former works on diffraction photodetector have shown the wavelength insensitivity of this effect.

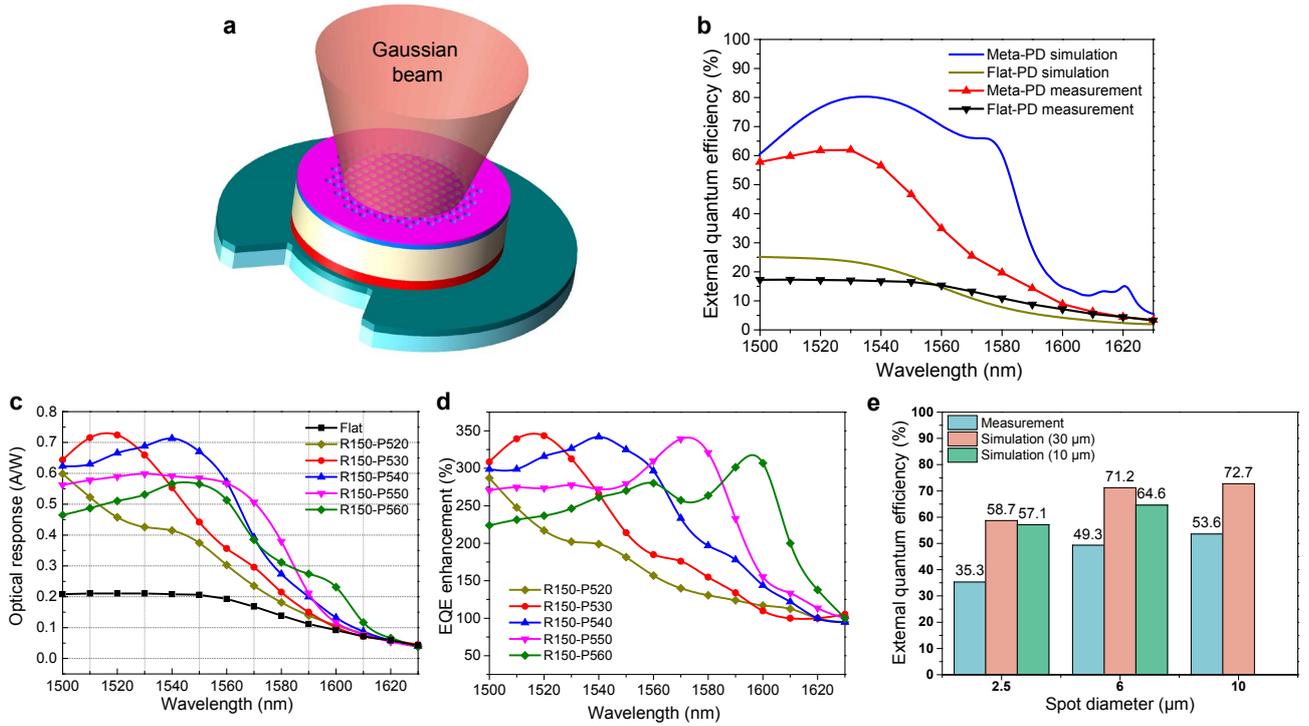

**Figure 4 | Enhanced external quantum efficiency of the full structure RM-Ge-PD. a**, Schematic of Gaussian beam incidence on the RM-Ge-PD, absence of the top SiO$_2$ layer is for a clear depiction. **b**, FDTD simulated and experimentally measured EQE spectra of the 15 μm-radius RM-Ge-PD with hole-array with diameter/period of 320 nm/540 nm. The measured EQE is above 62% at 1530 nm and 50% at 1550 nm, while the simulated one is as high as 80% around 1540 nm. Both results show strongly enhanced absorption compared with flat device, respectively. **c,d**, Measured responsivity spectra (**c**) and the corresponding EQE enhancement spectra (**d**) for RM-Ge-PD as the hole-array (hole radius: 150 nm) period increases from 520 to 560 nm, exhibiting high responsivity of more than 0.7A/W and enhancement peaks of more than 300% from 1500 to 1600 nm. The peak wavelength drift is almost linear with the variation of period. **e**, FDTD simulated and experimentally measured EQEs of RM-Ge-PD at 1550 nm with different incident spot diameter and mesa diameter. The diameter/period of the metasurface is 320 nm/540 nm. (Notice: the depth of the hole-array is 215 nm for all the simulated and fabricated devices in Fig. 4.)

In addition to a thin intrinsic layer, a small mesa is also required to obtain a high speed photodetector, as the junction capacitance can be severely decreased with a smaller size of mesa. The unique property of metasurface gives great convenience for the functional verification of photodetector with different size. As the operation region of the metasurface is intersection of the metasurface region and the incident spot. Fig. 4e exhibits the experimentally measured EQEs (@1550 nm) of the 15 μm-radius RM-Ge-PD with different incident spot sizes, along with the related simulation results. Tapered lensed fibers are used to obtain incident spot diameters of 10 μm, 6 μm and 2.5 μm in experiment, and the measured EQEs are 53.6%, 49.3% and 35.3%, respectively. The corresponding FDTD simulated absorption efficiencies of the 15 μm-radius RM-Ge-PD are 72.7%, 71.7%, and 58.7%. Considering the dissipation of carriers in the doped Ge layers, the experimental results are in good consistence with the simulation ones. The proximity between the results with spot size of 10 μm and 6 μm shows that the metasurface is quite tolerant with the operation size in our device. For dielectric metasurface with high Q-factor in non-absorption material, the performance is very sensitive to the operation size of metasurface, due to that the discontinuity at the edges brings extra loss. While in our case of metasurface with low Q-factor, the absorption loss in Ge is the main dissipation way of photon, the resonance of metasurface can be maintained within in a few cycle counts.

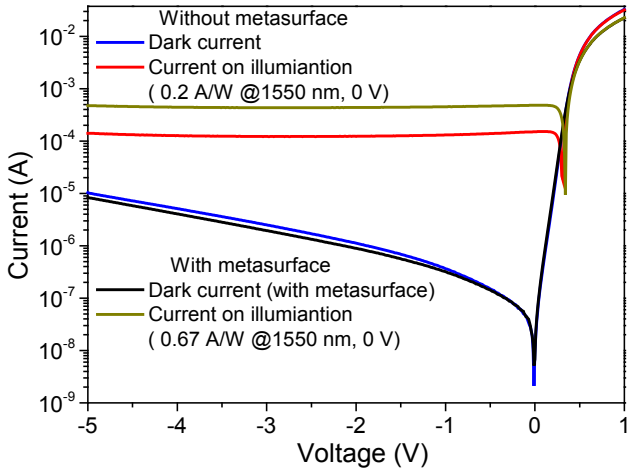

**Figure 5 |** The I-V characteristics of the PDs with and without resonant-metasurface. (period: 540 nm, radius: 150 nm).

The incident spot size of 2.5 μm gives an extreme situation that the resonance is barely maintained with too few cycle counts. Furthermore, the simulation of a 5 μm-radius RM-Ge-PD is implemented with incident spot diameters of 6 μm and 2.5 μm, to prove the feasibility of high speed device. Compared with the simulation results of 15 μm-radius RM-Ge-PD, the degeneration in performance is negligible. The small decrease in EQEs probably comes from the diffraction effect which needs a large lateral dimension to absorb the trapped photon.

## The electrical characteristics
### 1. Static Measurements

The fabricated photodetectors were biased at varying voltages, with and without light incidence, the results are shown in Fig. 5. The 15 μm-radius RM-Ge-PD exhibits dark currents as low as 5 nA and 320 nA at 0 V and 1 V, corresponding to dark current densities of 0.71 mA/cm$^2$ and 45.2 mA/cm$^2$, respectively. Compared with the results of flat PD tested in the same condition (dark currents of 2 nA and 360 nA, dark current densities of 0.28 mA/cm$^2$ and 51 mA/cm$^2$), it is clear that the shallow-etched holes do not increase the dark current of the PDs. As the hole-array is only etched to the upper part of the intrinsic layer, no extra leakage current can be introduced at the side walls of these holes (ref. 9). Weeny differences of dark currents between different PDs primarily come from the fabrication randomness. Different from former works of efficiency enhanced photodetectors with etched-through holes, our structure gives great convenience for the device fabrication, as no extra passivation process is required. A tunable semiconductor laser of 1 pm resolution is used for the responsivity measurement, with an incident optical power of

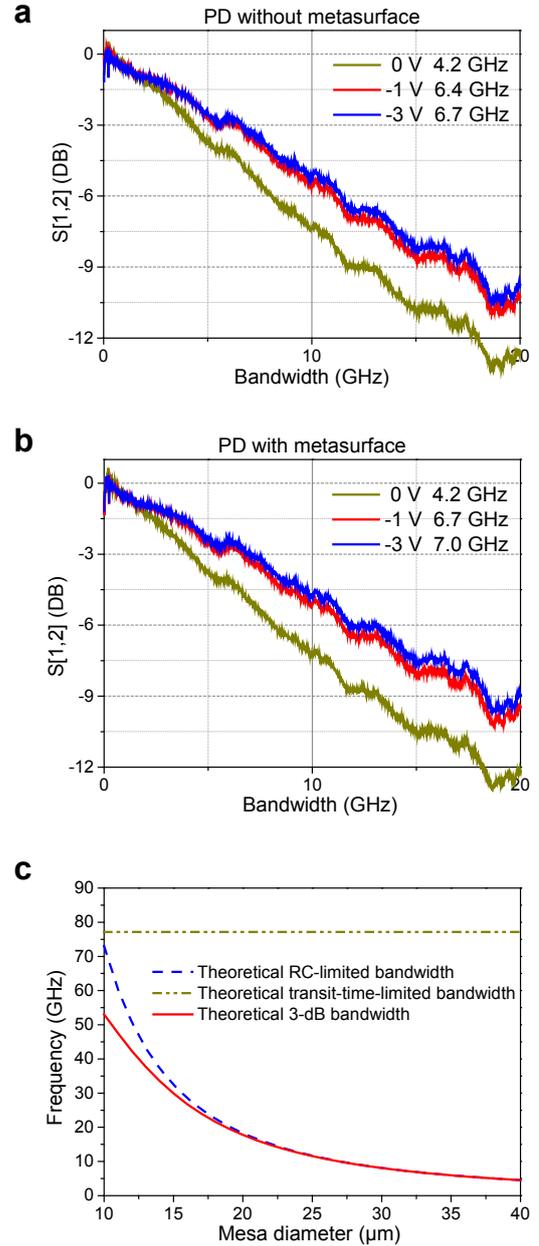

**Figure 6 | a**, Normalized frequency responses of the PD without holes at 1550 nm. **b**, Normalized frequency responses of the PD with holes at 1550 nm. **c**, Theoretical calculation about transit-time-limited bandwidth, RC-limited bandwidth, and 3 dB bandwidth of the PDs with different diameters, when thickness of the intrinsic layer is 350 nm.

0.75 mW at 1550 nm. Optical responsivities of the patterned and flat PDs are 0.67 A/W and 0.2 A/W at 0 V bias, corresponding to the EQEs of 53.6% and 16%, respectively. It is important to notice that the responsivity slightly decreases as the reverse bias increases, due to the Franz-Keldysh effect (FKE)[31]. For example, optical responsivity of the device without holes are 0.175 A/W and 0.16 A/W at -1 V and -3 V, respectively. The strong FKE indicates that the PDs can work at low bias voltage, which is meaningful for low power consumption receivers.

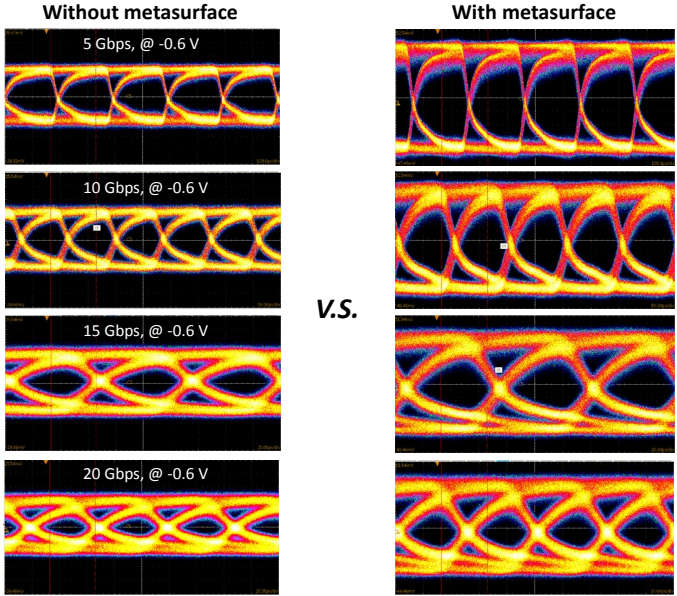

**Figure 7** | 5 Gbps，10 Gbps，15 Gbps，20 Gbps eye diagrams of the devices at a bias of -0.6 V.

## 2. Small-Signal Measurements

Frequency response of the device is measured by a vector network analyzer (VNA). The modulated light at 1550 nm is coupled to the device, and the electrical output is measured through a high speed RF probe. As shown in Figs. 6a and 6b, 3 dB bandwidths of the flat PD are 4.2 GHz, 6.4 GHz and 6.7 GHz at bias of 0 V, -1 V and -3 V. And 3 dB bandwidths of the patterned PD are 4.2 GHz, 6.7 GHz and 7 GHz at bias of 0 V, -1 V and -3V, respectively. From -1 V to -3 V, 3 dB bandwidths of the PDs change almost insignificantly, which indicates that the devices can work at high speed under low voltage. From the comparison between the two kinds of PDs, we can see that the introduced hole-array not only provides strongly enhanced optical absorption, but also reduces the junction capacitance by decreasing the junction area, leading to a better 3 dB bandwidth.

Theoretically, the bandwidth of a common p-i-n photodetector is mainly determined by the carrier transit-time-limited bandwidth ($f_T$) and resistor-capacitor bandwidth ($f_{RC}$) in the active region. $f_T$ and $f_{RC}$ can be approximated using the following equations[32]:

$$f_T = \frac{0.45 v_{sat}}{d_i}, f_{RC} = \frac{1}{2\pi(R_L + R_S)C}, f_{3dB} = \sqrt{\frac{1}{f_T^{-2} + f_{RC}^{-2}}}$$

Where $v_{sat}$ is the saturated velocity of Ge ($v_{sat} = 6 \times 10^6 \, cm/s$), $d_i$ is the thickness of intrinsic layer, $C$ is the capacitance of PD, $R_L$ is the load resistance ($R_L = 50 \, \Omega$), and $R_S$ is the series resistance. The theoretically calculated transit-time-limited bandwidth, RC-limited bandwidth, and the 3 dB bandwidth of the devices is shown in Fig. 6c. Specially, the calculated 3 dB bandwidth of our device with a 30 μm-diameter mesa is 8 GHz, which is in good consistent with the measured result. When thickness of the intrinsic layer is 350nm, transit-time-limited bandwidth is up to 77 GHz. The 3 dB bandwidth increases monotonously as the mesa size decreases, with a notably high value of more than 50 GHz at a mesa diameter of 10 μm.

## 3. Large-Signal Measurements

To further test the performance of the devices in high speed optical communication, an eye-diagram measurement was carried out using a similar experimental setup as the bandwidth measurement. An optical nonreturn-to-zero (NRZ) pseudorandom binary sequence (PRBS) signal generated by a commercial LiNbO$_3$ modulator at 1550 nm is delivered to the photodetector. Loaded with a low bias of -0.6 V through a bias-tee, the eye diagrams of the patterned and flat PDs at 5 Gbps, 10 Gbps, 15 Gbps and 20 Gbps are shown in Fig. 7. The clearly open eyes of the flat PD show high performance data reception at 5 Gbps, 10 Gbps and 15 Gbps. And the quality of the open eyes deteriorates a little bit when the signal rate reaches up to 20 Gbps, but the eyes still open clearly. Eye diagrams of the patterned PD show higher signal level under the same measurement condition, because of higher optical responsivity. It is worth noticing that the eye diagrams of the patterned PD become a little bit fuzzy at 15 Gbps and 20 Gbps, because space-charge effect causes the drift speed of carriers to slow down under large photocurrent. And it can be solved by decreasing the power of the incident light.

## Conclusion

We have demonstrated a normal incident Si-based RM-Ge-PD with an ultra-thin intrinsic Ge layer of 350 nm, which is the first experimental demonstration of Ge photodetector integrated with resonant metasurface. Over 50% external quantum efficiency for 1500-1550 nm wavelengths, and a peak value of 62% at 1530 nm can be achieved. This is the highest quantum efficiency achieved for a normal incident Ge photodetector with such thin intrinsic layer and wide operation spectrum. Despite the 15 μm-radius mesa size, the eye-diagram measurement shows normal operation at the signal rate of 20Gbps. Our work reveals a flexible way to strongly enhance the responsivity by trapping the incident photons in the collective resonance of the metasurface, and the functional spectrum can be widened by overlapping multiple resonant modes. Detailed comparison between experiment and simulation shows the feasibility of absorption strongly enhanced Ge photodetector with 5 μm-radius mesa size, corresponding to 3dB-bandwidth more than 50 GHz. Furthermore, the structure can also be implemented in the 1310 nm wavelength range with even better performance. With great potential to reduce the cost of packaging and heat generation, this enables the development of efficient high-speed Ge photodetector compatible for monolithic integration with CMOS electronics for intra- and inter-data centers communication